\documentclass[12pt]{article}
\usepackage{color}
\usepackage{scicite}
\usepackage{times}
\usepackage{graphicx,amsmath,amssymb}
\usepackage{graphics,color,epsfig}
\newenvironment{sciabstract}{%
\begin{quote} \bf}
{\end{quote}}

\newcounter{lastnote}
\newenvironment{scilastnote}{%
\setcounter{lastnote}{\value{enumiv}}%
\addtocounter{lastnote}{+1}%
\begin{list}%
{\arabic{lastnote}.}
{\setlength{\leftmargin}{.22in}}
{\setlength{\labelsep}{.5em}}}
{\end{list}}

\title{Heavy Fermions and Quantum Phase Transitions}

\author{Qimiao Si,$^{1,\ast} $ Frank Steglich$^{2,\ast}$
\\
\\
\normalsize{$^{1}$ Department of Physics \& Astronomy, Rice University,
Houston, TX 77005, USA}\\
\normalsize{$^{2}$ Max Planck Institute for Chemical Physics of
Solids, 01187~Dresden, Germany}\\
\\
\normalsize{$^\ast$To whom correspondence should be addressed; E-mails:}\\
\normalsize{qmsi@rice.edu; steglich@cpfs.mpg.de.}
}

\date{}
\begin{document}
\newcommand{\gc}{\color{green}}
\baselineskip24pt \maketitle

\begin{sciabstract}
%  \underline{One-sentence Summary:}~~
%\\
% \underline{Abstract:}~~
Quantum phase transitions arise in many-body systems due to 
competing interactions that promote rivaling ground states.
Recent years have seen the identification of continuous quantum
phase transitions, or quantum critical points, in a host of
antiferromagnetic heavy-fermion compounds. Studies of the interplay
between the 
%Kondo effect and magnetism have brought out
various effects have revealed
new classes of quantum critical points, and are uncovering 
a plethora of new quantum phases. At the same time,
quantum criticality has provided fresh insights into the electronic,
magnetic, and superconducting properties of  the heavy-fermion 
metals. We 
review 
these developments, discuss the open issues, 
and outline some directions for future research.
\end{sciabstract}
% In setting up this template for *Science* papers, we've used both
% the \section* command and the \paragraph* command for topical
% divisions.  Which you use will of course depend on the type of paper
% you're writing.  Review Articles tend to have displayed headings, for
% which \section* is more appropriate; Research Articles, when they have
% formal topical divisions at all, tend to signal them with bold text
% that runs into the paragraph, for which \paragraph* is the right
% choice.  Either way, use the asterisk (*) modifier, as shown, to

\newpage

Quantum mechanics not only governs the subatomic world, but also
dictates the organization of the microscopic particles in bulk
matter at low temperatures. The behavior is strikingly different
depending on the spin (the internal angular momentum) of the
constituent particles. Particles whose spin is an integer multiple
of ${\rm \hbar}$ are bosons. When cooled down to sufficiently low
temperatures, they will be described by the same wavefunction,
thereby forming a ``condensate''. Particles whose spin is
a half-integer of ${\rm \hbar}$, on the other hand, are fermions
satisfying the Pauli exclusion principle; no two particles can have
the same state. At absolute zero, they occupy the states with the
lowest energies, up to an energy referred to as the Fermi energy.
In the momentum space, this
defines a Fermi surface, enclosing a Fermi volume in which all the
states are occupied.

When the particle-particle interactions are included,  the behavior
of such quantum systems becomes even richer. These 
strongly correlated systems have taken the center stage in the field
of quantum matter over the past two decades \cite{S.I.Science-08}.
%In the electron world, h
High temperature superconductors, fractional
quantum Hall systems, 
colossal
magnetoresistive materials,
and magnetic heavy-fermion metals
are 
%but 
a few prominent examples. The
central question for all these systems is how the electrons are
organized, and, in particular,
whether there are 
principles that are universal
among the various classes of these strongly correlated materials.
One such principle, which has come  to the forefront in recent
years, is quantum criticality %\cite{Natphys-qpt08}.
\cite{PSS-qcnp10}.

A quantum critical point (QCP) arises when matter undergoes a continuous
transition from one phase to another at zero temperature. A non-thermal
control parameter, such as pressure, tunes the amount of zero-point
motion of the constituent particles. In other words, it controls
quantum-mechanical tunneling dictated by Heisenberg's 
uncertainty principle, thereby changing the degree of quantum fluctuations.
This is the analogue
of varying the thermal fluctuations in the case of temperature-driven
classical phase transitions, such as the melting of ice or the loss
of ferromagnetic order in iron.

Figure 1A illustrates the temperature-pressure phase diagram
observed in the heavy-fermion intermetallic compound CePd$_2$Si$_2$
\cite{Grosche-2001}.
At ambient pressure, CePd$_2$Si$_2$ orders into
an antiferromagnet, below the N\'{e}el temperature  $T_{\rm N}$ of about 
10 K.
Applying pressure reduces $T_{\rm N}$ monotonically, eventually
suppressing the antiferromagnetic order altogether and turning the
system into a paramagnetic metal. The putative critical pressure,
$p_{\rm c}$, is around 2.8 GPa, where an antiferromagnetic 
QCP is implicated. The QCP, however, is not
explicitly observed. Instead, a ``dome'' emerges at very low
temperatures in the vicinity of $p_{\rm c}$, under which the system
is a superconductor.

This phase diagram exemplifies a general point. It 
suggests that
antiferromagnetic quantum criticality 
can provide a mechanism for
superconductivity, an observation that may be of relevance to a
range of other strongly correlated systems such as high-$T_{\rm c}$
cuprates, organic superconductors and the recently discovered
high-$T_{\rm c}$ iron pnictides. The formation of new phases near
a QCP may be considered  the consequence of an
accumulation of entropy, which is a generic feature of any 
QCP 
%\cite{ZhuGarst.03} 
\cite{Gegenwart.08}
and has recently been observed
experimentally %\cite{kuech09,tokiw09,Rost.09}.
\cite{Gegenwart.08,tokiw09,Rost.09}.

A good example for such an antiferromagnetic QCP is the one observed
in the compound YbRh$_2$Si$_2$ \cite{Gegenwart.08}.
Here, the non-thermal
control parameter is a (small) magnetic field. Figure 1B brings out
another important new insight that has been gained from the studies
of heavy-fermion antiferromagnets. Accompanying the QCP at
zero-temperature is a finite parameter range at non-zero
temperatures in which the metallic state is anomalous
\cite{custe03,Friedemann09}.
Over this quantum critical regime, the electrical
resistivity is linear in temperature --
%. As we will describe in some detail later, this linearity is 
a telltale sign for an unusual
metallic state.
This non-Fermi liquid behavior \cite{Maple.94},
which goes beyond the standard theory of metals,
Fermi-liquid theory \cite{Hewson}, is another phenomenon 
that is broadly relevant to the physics of strongly correlated
systems \cite{Hussey.09,Grigera.01}.

Quantum criticality has been implicated to one degree or another in a host of
other heavy-fermion metals \cite{Gegenwart.08,stewa01,HvL-RMP}.
These include
%Among them are 
CeCu$_2$Si$_2$,
the first superconductor to be observed among heavy-fermion metals
\cite{Steglich79}, and CeRhIn$_5$ \cite{hegge00}
%, as shown in Figure 1c.
(Figure 1C).
Extensive 
theoretical studies have led to 
a novel type of quantum 
criticality \cite{Si-Nature,Colemanetal,senthil2004a,Pepin}.
More recently, a plethora of new phases have been uncovered in 
heavy-fermion metals near a QCP ({\it e.g.}, 
%Refs.~
in Ir-doped YbRh$_2$Si$_2$ \cite{Friedemann09}
and in $\beta$-YbAlB$_4$ \cite{Nakatsuji09}).
Together with the theoretical studies of the global phase diagram
of the heavy-fermion metals \cite{Si.06,PSS-qcnp10},
these developments open up an entirely new frontier
on 
the interplay between quantum criticality and novel phases.

%\section{INTRODUCTION}

%\subsection{Quantum criticality}
\section*{Quantum Phase Transitions}

Quantum phase transitions
result from the variation of quantum fluctuations.
Tuning a control parameter at absolute zero temperature 
tilts the balance among the
competing ground states associated with
conflicting
interactions
of quantum matter.

Heavy-fermion metals comprise a lattice of localized magnetic
moments and a band of conduction electrons \cite{Hewson}. The
exchange interaction between the local moments is primarily that
mediated by the conduction electrons, the familiar
Ruderman-Kittel-Kasuya-Yoshida (RKKY) interaction. This interaction
drives the local moments into an ordered pattern,
much like H$_2$O molecules are
condensed into an ordered arrangement in ice. The Kondo-exchange
interaction between the local moments and conduction electrons
introduces spin flips, a tunneling process enabled by quantum
mechanics. Correspondingly, increasing the Kondo interaction
amounts to enhancing quantum fluctuations, which eventually destroys
the magnetic order and yields a paramagnetic phase
\cite{Doniach,Varma76}.

The theory of classical phase transitions, formulated by Landau
\cite{Ma-book}, is based on the principle of spontaneous symmetry breaking. 
Consider CePd$_2$Si$_2$ at ambient pressure.
In the
paramagnetic phase, at $T>T_{\rm N}$,
the spins are free to rotate. Upon entering the
magnetically-ordered phase, this continuous spin-rotational symmetry
is spontaneously broken; the spins must choose preferred orientations.
In the
Landau formulation, this symmetry distinction is characterized by a
quantity called order parameter; in our case, this is the
staggered magnetization of the antiferromagnet. The order parameter
is nonzero in the magnetically-ordered phase, but
vanishes in the paramagnetic phase. The critical point 
arises when the phase transition is continuous, {\it i.e.}, when
the order parameter goes to zero smoothly. It 
is described in terms
of the spatial fluctuations of the order parameter. These fluctuations
occur over a characteristic length scale, which increases on approaching the
critical point.
% from the paramagnetic side. 
At the critical point,
the correlation length is infinite. Correspondingly, physical
properties are invariant under a mathematical operation which
dilates the lengths; in other words, they are scale invariant.

A straightforward generalization of the Landau paradigm to QCPs
gives rise to essentially the same
theoretical description \cite{Hertz}. Quantum mechanics introduces a
``time'' axis:
quantum states evolve in time. (For quantum systems in equilibrium, 
the relevant
quantum evolution is along an imaginary time of length
$\hbar/k_B T$.)
This introduces a time scale that accompanies the
divergent correlation-length scale. When the transition takes place
at a finite temperature $T_{\rm N}$, ${\rm \hbar/k_B}T_{\rm N}$
serves as the upper bound of the correlation time, and the ultimate
critical behavior is still determined by the fluctuations in space
only. When  $T_{\rm N}$ is driven to zero temperature,
however,
a divergent correlation time, $\xi_{\tau}$,
accompanies the divergent correlation
length, $\xi$, and both must be taken into account even
for equilibrium properties. Hence, the quantum critical fluctuations
of the order
parameter take place both in space and in time. 
The effective dimensionality of the fluctuations is $d+z$, where $d$
is the spatial dimensionality and, $z$, the dynamic exponent
defined in terms of the relationship $\xi_{\tau} \propto \xi^z$,
describes the number of effective spatial dimensions that the time
dimension corresponds to.

%In recent years, however, 
However, it has been appreciated that this Landau
paradigm can break down for QCPs.
Consider the effect of the Kondo-exchange coupling. 
In addition to destabilizing the magnetic order,
the Kondo interaction also introduces quantum coherence between the
local moments and conduction electrons. Indeed, inside the
paramagnetic phase, a process called Kondo screening takes place,
which leads to a qualitatively new ground state in which 
the local moments and conduction electrons are entangled.
Just as a continuous onset of magnetic order at
zero temperature introduces quantum fluctuations of the magnetic
order parameter, a critical 
onset 
of Kondo 
entanglement
also yields its own quantum critical
degrees of freedom. When
that happens, a new type of QCP 
ensues.

\section*{Kondo Effect and Heavy Fermions}

Historically, the Kondo screening effect
was introduced for dilute magnetic impurities in metallic
hosts \cite{Hewson}.
By the 1970s, the notion that the Kondo phenomenon
%effect
operates in a dense
periodic array of magnetic Ce ions in intermetallic compounds,
{\it e.g.} CeAl$_2$ \cite{bredl78}, was already in place. 
A characteristic scale, 
%below 
at which the Kondo screening initially sets in,
is  the Kondo temperature $T_{\rm K}$.

The list of heavy-fermion materials is 
long, and
they are typically compounds containing rare-earths or actinides
(including Yb, U, Np, in addition to Ce)
with partially-filled 4f- or 5f- orbitals. Their defining
characteristics is that the effective mass of the charge carriers
at the lowest accessible temperatures is hundreds of times
%or even thousands times of 
the bare electron mass.

Microscopically, 
heavy-fermion systems can be modeled as 
a lattice of localized f-electron moments
that are coupled to a band of conduction electrons.
In the early 1980s, the description of the Kondo 
%screening
effect in the
ground state of this Kondo lattice was formulated
\cite{Hewson}. The local moments
lose their identity 
by forming a many-body spin singlet with all
the conduction electrons,
leading to an entangled state (Fig.~2A).
The Kondo entanglement in the ground state makes the 
local moments, which are charge neutral to begin with, acquire 
the quantum numbers of the conduction electrons, namely spin-$\hbar/2$ 
and charge-$e$. Correspondingly, ``Kondo resonances'' appear as 
charge carriers, and they remember their localized-moment
origin by possessing a heavy mass. 
%Since 
As the Kondo resonances are 
part of the electronic-excitation spectrum, 
they must be accounted for in the Fermi surface, leading to the 
notion of a large Fermi surface (Fig.~2B) --
%. This is 
the picture of a heavy Fermi liquid.

The Kondo resonances can alternatively be thought
of as the remnants of the original f-electrons. They are
de-localized because the 4f- or 5f- wavefunction has 
a finite overlap with the ligand orbitals that form 
the conduction electrons. In other words, the f-electrons and 
conduction electrons are hybridized.

%\section{QUANTUM CRITICAL POINTS IN HEAVY FERMIONS}
\section*{Quantum Critical Points in Heavy Fermions}

%\subsection*{Two types of quantum critical points} 
\paragraph*{Two types of quantum critical points.} 
The Kondo singlet in the ground state of a heavy-fermion 
paramagnet represents an organized macroscopic pattern of the quantum 
many-body system (Fig.~2A). It endows 
the paramagnetic phase 
at zero temperature 
with a quantum order. 
This characterization of the phase goes beyond the Landau framework.
The Kondo-singlet state does not invoke any spontaneous breaking
of symmetry, as 
the spins can orient in arbitrary directions;
no Landau order parameter can be associated to the Kondo effect.
Two types of QCPs arise, depending on the behavior of the Kondo
singlet as we approach the QCP from the paramagnetic side.

When the Kondo singlet is still intact across the antiferromagnetic 
transition at zero temperature, the only critical degrees of freedom
are the fluctuations of the magnetic order parameter. In this case,
the antiferromagnetically-ordered phase in the immediate proximity to the 
QCP can be described in terms of a spin-density-wave (SDW) order of the 
heavy quasiparticles of the paramagnetic phase. The QCP is referred to
as of the SDW type, which is in the same class as that 
already considered by Hertz \cite{Hertz,Millis,Moriya,Continentino}.
On the other hand, when the Kondo singlet exists only
in the paramagnetic phase, the onset of magnetic order 
is accompanied by a breakdown of the Kondo effect.
The quantum criticality incorporates not only
the slow fluctuations of the antiferromagnetic order parameter, but also the 
emergent degrees of freedom
associated with the breakup of the Kondo singlet.
The corresponding transition is referred to as locally 
critical \cite{Si-Nature,Colemanetal}; the antiferromagnetic transition 
is accompanied by a localization of the f-electrons.

This distinction of the two types of QCPs can also be made in terms 
of energetics. The key quantity to consider is the
energy scale $E^\star$,
which dictates the breakup of the entangled Kondo singlet state
as the system moves from the heavy-Fermi-liquid side towards
the quantum critical regime. A reduction of the $E^\star$ scale on 
approaching the magnetic side is to be expected, as the development 
of antiferromagnetic
correlations among the local moments 
reduces 
the strength of the Kondo
singlet \cite{Si-Nature,Colemanetal,senthil2004a,Pepin}.
When $E^\star$ remains finite at the antiferromagnetic QCP, 
the Kondo singlet is still formed, and 
the quantum criticality falls in the universality class of the
SDW type.
When the $E^\star$ scale continuously goes to zero at the antiferromagnetic 
QCP, a critical Kondo breakdown accompanies the magnetic transition.
Notice that the $T_{\rm K}$ scale, where the Kondo screening initially
sets in, is always nonzero near the QCP, even when $E^\star$ approaches
zero. 

Figure 2 illustrates the consequence of the Kondo breakdown for
the change of the Fermi surface. When $E^{\star}$ is finite,
the Kondo-singlet ground state
supports Kondo resonances,
and the Fermi surface is large.
When the $E^{\star}$ scale becomes zero,
the ground state is no longer
a Kondo singlet and there are no fully-developed
Kondo resonances. Correspondingly, the Fermi surface is small,
incorporating only the conduction electrons.

In the Kondo-screened paramagnetic phase (Fig.~2A), the large
Fermi surface is where the heavy quasiparticles are located 
in the momentum space (Figure
2B). As usual, such sharply defined Fermi surfaces occur below an
effective Fermi temperature, $T_{\rm FL}$. Below this temperature,
standard
Fermi-liquid properties, such as 
the inverse 
quasiparticle lifetime and the electrical resistivity
being quadratically dependent on temperature, take place. 

In the Kondo-destroyed antiferromagnetic phase (Fig.~2C), 
there is no Kondo singlet in the ground state and,
correspondingly, static Kondo screening is absent.
Kondo screening still operates dynamically,
leading to an enhancement of the mass of the quasiparticles. 
The quasiparticles still have a Fermi-liquid form at low temperatures.
In contrast to the case of the Kondo-singlet ground state,
these quasiparticles are adiabatically connected to the ordinary
conduction electrons and are located at the 
small Fermi surface (Fig.~2D).

The large number of available compounds 
is a key advantage in the study of quantum critical heavy-fermion
systems.
At the same time, it raises an important question: can we classify 
the quantum critical behavior observed in these heavy-fermion compounds?
Below, we summarize the evidence for such a classification
in the systems that have been most extensively studied in the
present context.

%\subsection*{Quantum critical point of the spin-density-wave type}
\paragraph*{Quantum critical point of the spin-density-wave type.}
The phase diagram
for CePd$_2$Si$_2$ (Fig.~1A) \cite{Grosche-2001}
is reminiscent of
theoretical discussions of unconventional superconductivity
near an SDW instability.
Unfortunately, because of the high pressure
necessary to access the QCP in this compound, 
it has not yet been possible to study either
the 
order
or
the fluctuation spectrum near the QCP.
CeCu$_2$Si$_2$ is an ideal system for such an investigation
as,
here, heavy-fermion superconductivity forms 
in the vicinity of an antiferromagnetic QCP at ambient/low
pressure.
Neutron diffractometry
revealed the antiferromagnetically-ordered 
state to be an incommensurate SDW with small
ordered moment 
(about $0.1 \mu_{\rm B}$/Ce), due to
the nesting of the renormalized Fermi surface \cite{stock04}.
Inelastic 
neutron-scattering studies on paramagnetic CeCu$_2$Si$_2$
have 
identified
fluctuations
close to the incommensurate ordering wavevector of the nearby SDW
and have shown that such fluctuations play a dominant role in 
driving superconducting pairing \cite{stock08},
%\cite{stock10},
confirming earlier theoretical predictions.

Another compound 
is CeNi$_2$Ge$_2$, 
where
the magnetic instability
may be achieved by slight volume expansion.
The critical
Gr\"{u}neisen ratio
in this system
diverges as $T^{-1}$,
%\cite{kuech09},
which lends 
support for a nearby SDW QCP \cite{Gegenwart.08}.
%\cite{ZhuGarst.03}.

There are also a few examples of magnetic quantum phase transitions
induced by alloying which appear to fall in the category of 
the SDW QCP.
In Ce$_{1-x}$La$_x$Ru$_2$Si$_2$, for instance, 
recent 
inelastic neutron-scattering 
experiments have provided 
such evidence near its critical
concentration $x_{\rm c} \approx 0.075$ \cite{Knafo09}.

%\subsection*{Quantum critical points involving a Kondo breakdown}
\paragraph*{Quantum critical point involving a Kondo breakdown.}
As shown in Fig.~3A, inelastic neutron-scattering 
experiments 
on the
quantum critical material CeCu$_{5.9}$Au$_{0.1}$ revealed an energy
over temperature ($E/T$) scaling \cite{Aronson.95}
of the dynamical susceptibility, with a
fractional exponent \cite{Schroder}. 
The same critical exponent is
found to govern the magnetic susceptibility at wavevectors far away
from the antiferromagnetic wavevector.
These features are incompatible with the predictions of the SDW
theory \cite{Hertz,Millis,Moriya,Continentino} and have 
provided the initial motivation for the development of 
local quantum criticality \cite{Si-Nature}.
%Since
As such a 
QCP
involves a breakdown of the Kondo effect,
it must be manifested in the charge carriers 
and their Fermi surfaces as well. 

Direct measurements of Fermi surfaces are typically done using the
angle-resolved photoemission (ARPES) spectroscopy. In spite of
impressive recent developments,
ARPES still does not have the 
resolution to study heavy-fermion metals in the required sub-Kelvin
low-temperature range. The other well-established means to probe
Fermi surfaces is the de Haas-van Alphen (dHvA) technique which,
however, requires a large magnetic field of several Teslas. A rare
opportunity arises in CeRhIn$_5$, where a magnetic field of about 10 Tesla
is, in fact, needed to suppress superconductivity and expose a
quantum phase transition (Fig.~1C). From dHvA
measurements performed in the field range 10-17 T, a pronounced
jump in the Fermi surface was 
seen in CeRhIn$_5$ at the critical pressure of 2.3 GPa \cite{shishido2005}
(Fig.~3B). This, together with the observation of a seemingly
diverging cyclotron mass of the heavy charge carriers, is commonly
considered as evidence for a Kondo-breakdown QCP \cite{park-nature06}.
We caution that, for CeRhIn$_5$,
this issue remains to be fully settled; an alternative
explanation based on 
a change of the valence state 
of the Ce ions has also been made \cite{watan09}.

The heavy-fermion 
%compound 
metal YbRh$_2$Si$_2$ has provided an
opportunity to probe 
the electronic properties near an antiferromagnetic QCP involving 
a breakdown of the Kondo effect.
As mentioned earlier, the very weak antiferromagnetic 
order of YbRh$_2$Si$_2$ is
suppressed by a small magnetic field, giving way to non-Fermi liquid
behavior \cite{custe03}. Furthermore,
the magnetic field
induces a substantial change of the isothermal Hall coefficient. The
latter has been shown to probe, at low temperatures,
the
properties of the Fermi surface
\cite{paschen2004}. A new 
temperature
scale,
$T^\star(B)$,
was
identified
in the $T-B$ phase diagram of YbRh$_2$Si$_2$
%, {\it cf.} Figure 1b;
(Fig.~1B);
across this scale, the isothermal Hall coefficient exhibits a
crossover as a function of the applied
magnetic field.
This crossover sharpens upon cooling.
Extrapolation to $T=0$ suggests
an abrupt change of the Fermi surface
at the critical magnetic field $B_{\rm N}$,
the 
field where
$T_{\rm N}$ approaches zero \cite{paschen2004}.
Further evidence for the inferred change of Fermi surface has come
from thermotransport measurements \cite{Hartmann09}.
Across the $T^\star$ line, the low-temperature thermopower shows
a sign change, suggesting an evolution between hole-like and
electron-like Fermi surfaces as illustrated in Fig.~2B,D.

%Subsequent
Further 
thermodynamic and transport investigations confirmed $T^\star(B)$ to be
an intrinsic 
energy scale which vanishes at the antiferromangetic QCP (Fig.~4A)
\cite{Gegenwart2007}. The $T^\star$ scale is distinct from the
Fermi-liquid scale, $T_{\rm FL}$, below which a $T^2$ resistivity 
is observed (Figure 4A).
These properties are naturally interpreted 
as signatures of a breakdown of the Kondo effect at the QCP,
with the Fermi surface being large at $B> B_{\rm N}$ (Fig.~2A,B)
and being small at $B < B_{\rm N}$ (Fig.~2C,D);
$T^\star$ refers then to
the temperature scale accompanying 
the Kondo-breakdown energy scale $E^\star$ introduced earlier.
It is worth noting that the $E^\star$ scale is distinct from 
the aforementioned $T_{\rm K}$ scale, which serves as the upper
cut-off of the quantum-critical scaling regime and should therefore 
remain finite near the QCP. 
For instance, at the critical concentration of CeCu$_{6-x}$Au$_{x}$,
$T_{\rm K}$ has been observed in photoemission spectroscopy
to be 
%remains 
non-zero \cite{Klein08} even though $E^\star$ is expected to vanish.

A recent
thorough study of the Hall crossover on YbRh$_2$Si$_2$ single
crystals of substantially improved quality showed unequivocally that
the width of the crossover at $T^\star(B)$ is strictly proportional to
temperature, {\it cf.}
Figure 4b \cite{friedemann_hall}.
This indicates that the energy over temperature
scaling also operates in this compound. Furthermore, it provides evidence
that the Kondo-breakdown effect 
indeed
underlies such quantum
critical scaling
\cite{friedemann_hall}.

%\section{GLOBAL PHASE DIAGRAM}
\section*{Global Phase Diagram}

The fact that, in YbRh$_2$Si$_2$, the multiple lines defining the
Kondo-breakdown scale $T^\star$, the Fermi-liquid scale $T_{\rm FL}$,
and the N\'{e}el-temperature scale $T_{\rm N}$ all converge at the
same magnetic field in the zero-temperature limit raises the
question of what happens when some additional control parameter is
varied. This global phase diagram has recently been explored by
introducing chemical pressure to YbRh$_2$Si$_2$ \cite{Friedemann09}.
The antiferromagnetic order is stabilized/weakened by volume
compression/expansion
(Fig.~4C-E),
in accordance
with the well-established fact that pressure 
reinforces magnetism in
Yb-based intermetallics.
Unexpectedly,
however, the $T^\star(B)$ line is only weakly
dependent on chemical pressure. Under volume compression 
(3\% Co-doping) the antiferromagnetic QCP occurs at
a field substantially higher than $B^\star$ at which
$T^\star\rightarrow 0$ (Fig.~4E). In this situation, $T^\star$ is finite
at the antiferromagnetic QCP.
One therefore expects that the
SDW description will apply, and this 
is indeed 
observed \cite{Friedemann09}.
Under a small volume expansion (2.5\% Ir-doping),
$B_{\rm N}$ and $B^\star$ continue to coincide within the experimental
accuracy (Fig.~4D). With a large volume expansion (17\% Ir-doping),
on the other hand, $B_{\rm N}$ has vanished but
$B^\star$ remains finite (Fig.~4C).
This opens up a range of magnetic field in which not only
any magnetic ordering is absent, but also the 
%Kondo-screening
Kondo-breakdown
scale vanishes,
suggesting
a small Fermi surface.
Hydrostatic-pressure experiments \cite{Tokiwa09} on
undoped YbRh$_2$Si$_2$
give 
 results 
comparable 
to those of the Co-doped materials
with a similar average unit-cell volume,
%chemical pressure,
indicating 
that the crossing of $T_{\rm
N}(B)$ and $T^\star (B)$
as observed ~\cite{Friedemann09}
for 7\% Co-doped YbRh$_2$Si$_2$
originates from the alloying-induced volume compression rather than
disorder.

The results can be summarized in the global phase diagram shown in
Fig.~4F. The transition from the small-Fermi-surface
antiferromagnet to the heavy-Fermi-liquid state has three types.
It may go through a large-Fermi-surface antiferromagnet, as in
the Co-doped cases. 
The transition 
can also occur directly,
as in the pure and 2.5\%-Ir-doped compounds.
Finally, it may go through a small-Fermi-surface paramagnetic phase
as in the case of the 6\% Ir-doped YbRh$_2$Si$_2$ \cite{Friedemann09}.
In this phase, the electrical resistivity shows a
quasi-linear temperature dependence
\cite{Friedemann09}.

Theoretically, two kinds of antiferromagnet to
% Kondo-screened 
heavy-Fermi-liquid transitions were already considered in the previous section.
One way to connect them is to invoke a $T=0$ global phase 
diagram \cite{Si.06}, spanned by two parameters associated with 
two types of quantum fluctuations.
One parameter, $J_K$, describes the Kondo coupling between
the conduction electrons and the local moments; increasing 
$J_K$ 
enhances 
the ability of the conduction electrons to 
screen the local moments and thereby reduces the magnetic order.
The other parameter, $G$, is associated with
the interactions among the local moments and 
refers to, for instance, the degree of geometric frustration \cite{Balents.10}
or simply the dimensionality \cite{Si-Nature,Matsuda.10};
raising the parameter $G$ boosts the inherent quantum fluctuations
of the local-moment system and, correspondingly, weakens the magnetism.
In the two-parameter global phase diagram
of Ref.~\cite{Si.06}, 
each kind of transition appears as a line of critical points:
one line is associated with local quantum criticality,
with the breakdown of the Kondo effect occurring 
at the antiferromagnetic-ordering
transition; the other one is associated with SDW quantum criticality,
in which case the Kondo breakdown can only take place 
inside the antiferromagnetically-ordered region.
This is consistent with Fig.~4F, in which $B_{\rm N}$ and $B^\star$
coincide for a finite range of small Ir-concentrations.
The extension of this global phase diagram is currently being 
actively pursued theoretically \cite{PSS-qcnp10}.
When the
quantum fluctuations among the local moments are even stronger,
a possibility exists 
for a paramagnetic phase with a suppressed Kondo 
entanglement and 
a concomitant small Fermi surface; this can be compared 
with the region 
highlighted by the question marks in Fig.~4F.
This phase could 
be a spin liquid, or could be an ordered state (such as 
a spin-Peierls phase) which preserves the spin-rotational
invariance. Understanding the nature of the novel phase 
represents an intriguing problem worthy of further studies, 
both theoretically and experimentally.

Other heavy-fermion systems may also be discussed in this two-parameter
global phase diagram. The zero-temperature transition in
CeCu$_{6-x}$Au$_x$ as a function of doping or
pressure can be 
described in terms of 
local quantum criticality.
As a function of magnetic
field, for both CeCu$_{6-x}$Au$_x$ \cite{Stockert_CeCuAu_field} and
CeIn$_3$ \cite{sebas09}, the Kondo breakdown seems to take place
inside the antiferromagnetic part of the phase diagram.
It will be instructive to see whether other heavy-fermion 
materials can be used to map the global phase diagram and,
in particular, display a paramagnetic non-Fermi liquid 
phase near a Kondo-breakdown QCP.

%\section{CONCLUSIONS AND OUTLOOK}
\section*{Conclusions and Outlook}

Studies in the last decade have firmly established the existence of
quantum critical points in heavy-fermion metals. These transitions 
arise from the suppression of
long-range antiferromagnetic ordering by tuning pressure, chemical
composition or magnetic field. An important property of
quantum critical points is the accumulation of entropy. Correspondingly, the
Gr\"{u}neisen ratio or the magnetocaloric effect diverges,
which
serves as an
important thermodynamic characterization of the
quantum critical points.

Two types of 
quantum critical points 
have been developed 
for antiferromagnetic heavy-fermion systems.
When a breakdown of the Kondo entanglement occurs inside the
\linebreak
antiferromagnetically-ordered phase, the quantum critical point 
has the standard spin-density-wave form which conforms to Landau's 
paradigm of order-parameter fluctuations.
When such a Kondo breakdown happens at the onset of
antiferromagnetism, a new class of quantum critical point 
arises. Evidence for this local quantum criticality 
has come from the quantum-dynamical
scaling and mass divergence 
in CeCu$_{\rm 6-x}$Au$_{\rm x}$ and YbRh$_2$Si$_2$,
the multiple energy scales observed in
YbRh$_2$Si$_2$, and the jump of the 
Fermi surface in YbRh$_2$Si$_2$ and CeRhIn$_5$.

A strong case has been made that, in CeCu$_2$Si$_2$, the critical
fluctuations of a spin-density-wave quantum critical point 
promote unconventional superconductivity.
It is likely that the
superconductivity in CePd$_2$Si$_2$ has a similar origin. Whether
the Kondo-breakdown local quantum critical points also favor
superconductivity is less clear. CeRhIn$_5$ under pressure and
$\beta$-YbAlB$_4$ could be examples in this category, although the
nature of quantum critical points 
in these systems remains to be firmly established.

More recent studies have focused 
attention, both experimentally
and theoretically,
on the global phase
diagram of antiferromagnetic heavy-fermion metals. Tantalizing
evidence has emerged for a non-Fermi liquid phase without any
magnetic ordering and with suppressed Kondo entanglement.
Whether such a state can, in fact, arise within the Kondo-lattice 
model is an intriguing
open theoretical question. 
In the process of addressing such issues,
it is becoming clear that quantum fluctuations
in heavy-fermion systems can be tuned in more ways
than one. Different phases and quantum critical points
may 
arise 
when a magnetic disordering is induced 
by the Kondo coupling between the local moments 
and conduction electrons, or when it is caused by reduced dimensionality 
and/or magnetic frustration.

Theoretically, an important notion which has emerged 
from studies in heavy-fermion systems is that
quantum criticality can go beyond the Landau paradigm 
of fluctuations 
in an order parameter associated with 
a spontaneous symmetry breaking.
This notion has impacted 
the developments 
on quantum criticality in other systems, including
insulating 
magnets.
More generally, quantum criticality in heavy-fermion metals 
epitomizes the richness and complexity of continuous quantum
phase transitions
compared to their classical counterparts. New theoretical methods
are needed to study strongly coupled quantum critical systems. One
promising new route is
provided by an approach based on quantum gravity \cite{Klebanov09}.
Using a charged black hole in a weakly-curved spacetime 
to model a finite density of
electrons, 
this approach has provided a tantalizing
symmetry reason for
some fermionic spectral quantities to display 
an anomalous frequency dependence when its momentum dependence 
is smooth. Whether 
a related symmetry principle underlies 
the dynamical scaling of the spin response at the Kondo-breakdown
local quantum criticality is an intriguing issue for future studies.

The insights gained from these studies on the well-defined 
quantum critical points in
various heavy-fermion metals have implications for 
other members of this class of materials as well as for 
other classes of strongly
correlated electronic systems. For example, an outstanding issue
is the nature of the hidden-order phase in 
the heavy-fermion compound URu$_2$Si$_2$ \cite{Davis.10,Yazdani.10}.
This phase is in proximity to some low-temperature magnetically-ordered
phases, raising the question of the role of quantum phase 
transitions in this 
exciting 
system.
In the cuprates, Fermi-surface
evolution as a function of doping has also been playing a prominent
role in recent years. In light of the discussions on the possible
role of doping-induced quantum critical points, it appears likely
that some of the physics discussed for heavy-fermion quantum
criticality also comes into play in the cuprates \cite{Taillefer.07}.
For the iron pnictides, the magnetic/superconducting phase
diagram has also been observed
to show a striking resemblance to Figure 1A.
The interplay between magnetic quantum criticality, electronic 
localization, and unconventional superconductivity, which has featured
so prominently in the systems considered here,
is likely pertinent to heavy-fermion metals in general as well
as other classes of correlated-electron materials, including 
the iron pnictides and organic charge-transfer salts.
Finally, quantum phase transitions are also being discussed
in broader settings, such as 
ultra-cold atomic gases
and quark matter. 
It is conceivable that issues related to our
discussion here will come into play in those systems as well.

%%\bibliography{review_complete.bib}
%\bibliography{review_complete}
%\bibliographystyle{Science}

%\begin{thebibliography}{99}
%
%\end{thebibliography}

%%%%%%%%%%%%%%%%%%%%%%%%%%% Acknowledgement%%%%%%%%%%%%%%%%%%%%%%%%%%%%%
% Following is a new environment, {scilastnote}, that's defined in the
% preamble and that allows authors to add a reference at the end of the
% list that's not signaled in the text; such references are used in
% *Science* for acknowledgments of funding, help, etc.

%%%%%%%%%%%%%%%%%%%%%%%%%%%%%%%%%%%%%%%%%%%%%%%%%%%%%%%%%%%%%%%%%%%%%%%%

\begin{scilastnote}
\item We would like to thank E. Abrahams, M. Brando, P. Coleman, 
S. Friedemann, P. Gegenwart, C. Geibel, 
F. M. Grosche, S. Kirchner, T. Park, J. Pixley, 
O. Stockert, J. D. Thompson, S. Wirth, and S. Yamamoto
for useful discussions. This work has been supported by 
NSF and the Robert A. Welch Foundation
Grant No. C-1411 (Q.~S.) and by the DFG Research Unit 960 
``Quantum Phase Transitions'' (F.~S.).
\end{scilastnote}

% For your review copy (i.e., the file you initially send in for
% evaluation), you can use the {figure} environment and the
% \includegraphics command to stream your figures into the text, placing
% all figures at the end.  For the final, revised manuscript for
% acceptance and production, however, PostScript or other graphics
% should not be streamed into your compiled file.  Instead, set
% captions as simple paragraphs (with a \noindent tag), setting them
% off from the rest of the text with a \clearpage as shown  below, and
% submit figures as separate files according to the Art Department's
% instructions.

\clearpage

%%%%%%%%%%%%%%%%%%%%%%%Figure Captions%%%%%%%%%%%%%%%%%%%%%%%%%%%%%%%%%%%%%%%%
\begin{figure}[h!]
\caption{Quantum phase transitions in heavy-fermion metals. (A)
Suppression of antiferromagnetic order by pressure in
CePd$_2$Si$_2$. $T_{\rm N}$ is the N\'{e}el transition temperature,
and the corresponding antiferromagnetic order is illustrated in the inset.
At the boundary of the antiferromagnetism, a phase of unconventional
superconductivity arises. 
$T_{\rm c}$ corresponds to the superconducting transition temperature.
(From Ref.~\cite{Grosche-2001}.)
(B) Field-induced quantum phase transition
in YbRh$_2$Si$_2$. The blue regions label the Fermi-liquid behavior
observed by measurements of electrical resistivity and other
transport and thermodynamic properties; they correspond to $T<T_{\rm
N}$ at $B<B_{\rm N}$, and $T<T_{\rm FL}$ at $B>B_{\rm N}$, where
$B_{\rm N}$ is the critical field at $T=0$. The orange region describes
non-Fermi liquid behavior that is anchored by the quantum critical
point at $B=B_{\rm N}$. (From Ref.~\cite{custe03}.)
The $T^\star$ line delineates crossover behavior
associated with the destruction of the Kondo effect, as described in
the main text. (From Ref.~\cite{Friedemann09}.)
(C) The pressure-field phase diagram at 
%$T=0$ 
the lowest measured temperature ($T=0.5$ K)
in CeRhIn$_5$. The antiferromagnetic order, denoted by MO,
at ambient pressure gives
way to superconductivity, specified by SC,
at higher pressures. 
At $B=0$, the antiferromagnetic order goes away when the pressure 
exceeds $P_1$. When the magnetic
field exceeds just enough to suppress superconductivity, 
the system is antiferromagnetically ordered at lower pressures, 
$P < P_2$, but yields a non-magnetic phase at higher pressures,
$P > P_2$. The hatched line refers to the transition at $P_2$, 
between these two phases. 
(From Ref.~\cite{park-nature06}.)
} \label{FIG1}
\end{figure}

\begin{figure}[h!]
\caption{Kondo entanglement and its breakdown in heavy-fermion metals. 
(A) Kondo-singlet ground state in a paramagnetic phase, giving rise
to a heavy Fermi liquid. The blobs with orange arrows mark the mobile 
conduction electrons, while the thick black arrows denote localized 
magnetic moments. The purple profile describes the Kondo singlet in the 
ground state. (B) The Kondo singlet in the ground state
gives rise to Kondo resonances,
which must be incorporated into the Fermi volume. Correspondingly, 
the Fermi surface is large,
with a volume that is proportional to $1+x$, 
where $1$ and $x$, respectively, refer to the number of local moments and 
conduction electrons per unit cell.
An SDW refers to an antiferromagnetic order which develops from
a Fermi-surface instability of these quasiparticles.
(C) Kondo breakdown in an antiferromagnetic phase. The local moments 
arrange into an antiferromagnetic order among themselves, and they do
not form static Kondo singlets with the conduction electrons.
(D) Kondo resonances do not form in the absence of static Kondo
screening. Correspondingly, the Fermi surface is small, 
enclosing a volume in the paramagnetic zone that is proportional to $x$.
Dynamical Kondo screening, however, still operates, giving rise to 
an enhancement of the quasiparticle mass near the small Fermi surface.
} \label{FIG2}
\end{figure}

\begin{figure}[h!]
\caption{Quantum critical properties of CeCu$_{6-x}$Au$_x$ and
CeRhIn$_5$. (A) Quantum-dynamical $E/T$ scaling of the inelastic
neutron-scattering 
cross section $S$ in CeCu$_{5.9}$Au$_{0.1}$. The measurements
were performed at the antiferromagnetic wavevectors (where $S$ is
maximized), and the scaling collapse is constructed in the form of
$T^{\alpha} S$ as a function $E/T$. The temperature and 
energy exponent is fractional: $\alpha=0.75$. The different 
symbols represent data
taken in different spectrometers at the different peak wavevectors.
Inset: The inverse of the bulk magnetic
susceptibility, $1/\chi({\bf q}=0) \equiv H/M$, and that of the
static susceptibility at other wavevectors derived from 
the dynamical spin susceptibility through the Kramers-Kronig
relation. The symbols not specified
in the legend correspond to other parts of the Brillouin zone.
(From Ref.~\cite{Schroder}.) (B) Several dHvA frequencies as a function
of pressure in CeRhIn$_5$. The applied magnetic field ranges between
10 T and 17 T.
$P_1$ and $P_2$ have the same meaning as in Fig.~1C.
The symbols denote different branches of the Fermi surface.
(From Ref.~\cite{shishido2005}.)
 } \label{FIG3}
\end{figure}

\begin{figure}[h!]
\caption{Quantum criticality and global phase diagram 
in pure and doped YbRh$_2$Si$_2$. 
(A) Multiple energy scales in pure YbRh$_2$Si$_2$. $T^\star$ is extracted
from isothermal crossovers in the Hall effect and thermodynamic
properties, which is interpreted in terms of a Kondo breakdown.
$T_{\rm FL}$ is the scale below which Fermi-liquid
properties occur.
Both crossover lines merge with the line that specifies the magnetic
phase boundary $T_{\rm N}$ in the zero-temperature limit, at $B_{\rm
N}$. 
(From Ref.~\cite{Gegenwart2007}.)
(B) Full width at half maximum (FWHM) of the crossover in the Hall coefficient
of a high-quality single crystal ($RRR = 120$).
It extrapolates to zero in the $T=0$ limit,
implying a jump of the Hall coefficient and other properties. It is
proportional to temperature, suggesting a quantum-dynamical $E/T$
scaling. (From Ref.~\cite{friedemann_hall}.)
C-E) $T^\star(B)$ and $T_{\rm N}(B)$ lines for Co- and Ir-doped
YbRh$_2$Si$_2$, determined via AC susceptibility measurements
\cite{Friedemann09}.
Data for the 7\% Co-doped 
YbRh$_2$Si$_2$ show an intersection of the 
two lines~\cite{Friedemann09}.
F) The $T=0$ phase diagram, doping-concentration
versus magnetic field, for Yb(Rh$_{1-x}$M$_x$)$_2$Si$_2$, M=Co,Ir,
from Ref.~\cite{Friedemann09}.} \label{FIG4}
\end{figure}

%\end{document}

\newpage
%%%%%%%%%%%%%%%%%%%%%%%%%%%%%%%%%%%%%%%%%%%%%%%%%%%%%%%%%%%%%%%%%%%%%%%%

%%\vspace{0.5cm}
%\begin{thebibliography}{99}
%
%\end{thebibliography}

%\begin{scilastnote}
%
%\end{scilastnote}

%\vskip 0.5 cm
%
%\noindent{\bf Supporting Online Material}
%
%\noindent{www.sciencemag.org}
%
%\noindent{Supporting online text}
%
%\noindent{Figs.~S1,S2}
%

\newpage
\begin{figure}[ht]
%%%\centerline{\includegraphics[width=0.8\linewidth]{Figure1_version10.eps}}
%\centerline{\includegraphics[width=0.6\linewidth]{Fig1b-figure_NP_new.eps}}
\centerline{\includegraphics[width=0.55\linewidth]{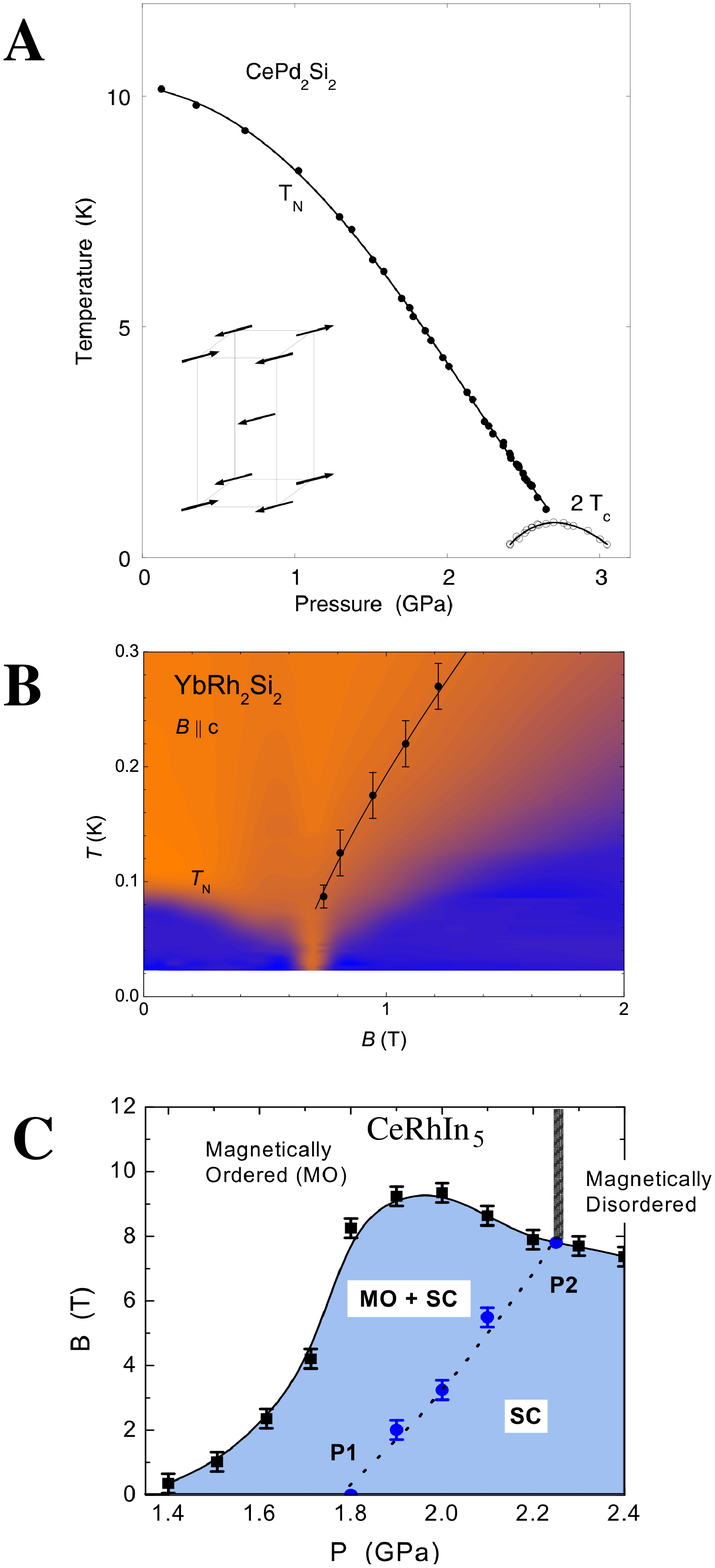}}
\centerline{} \centerline{} \centerline{\Large Figure~1}
\end{figure}

\newpage

\begin{figure}[h!]
%%%\centerline{\includegraphics[width=0.8\linewidth]{Figure2_version9.eps}}
%%\centerline{\includegraphics[width=0.6\linewidth]{figure2a.eps}}
%%\centerline{\includegraphics[width=0.6\linewidth]{figure2b.eps}}
%\centerline{\psfig{figure=figure2a.eps,width=0.4\linewidth,angle=0}
%\hskip 0.2cm
%\psfig{figure=figure2b.eps,width=0.4\linewidth}}
%\vskip 0.5 cm
%%\centerline{\includegraphics[width=0.6\linewidth]{figure2c.eps}}
%%\centerline{\includegraphics[width=0.6\linewidth]{figure2d.eps}}
%\centerline{\psfig{figure=figure2c.eps,width=0.4\linewidth,angle=0}
%\hskip 0.2cm
%\psfig{figure=figure2d.eps,width=0.4\linewidth}}
\centerline{\includegraphics[width=1.0\linewidth]{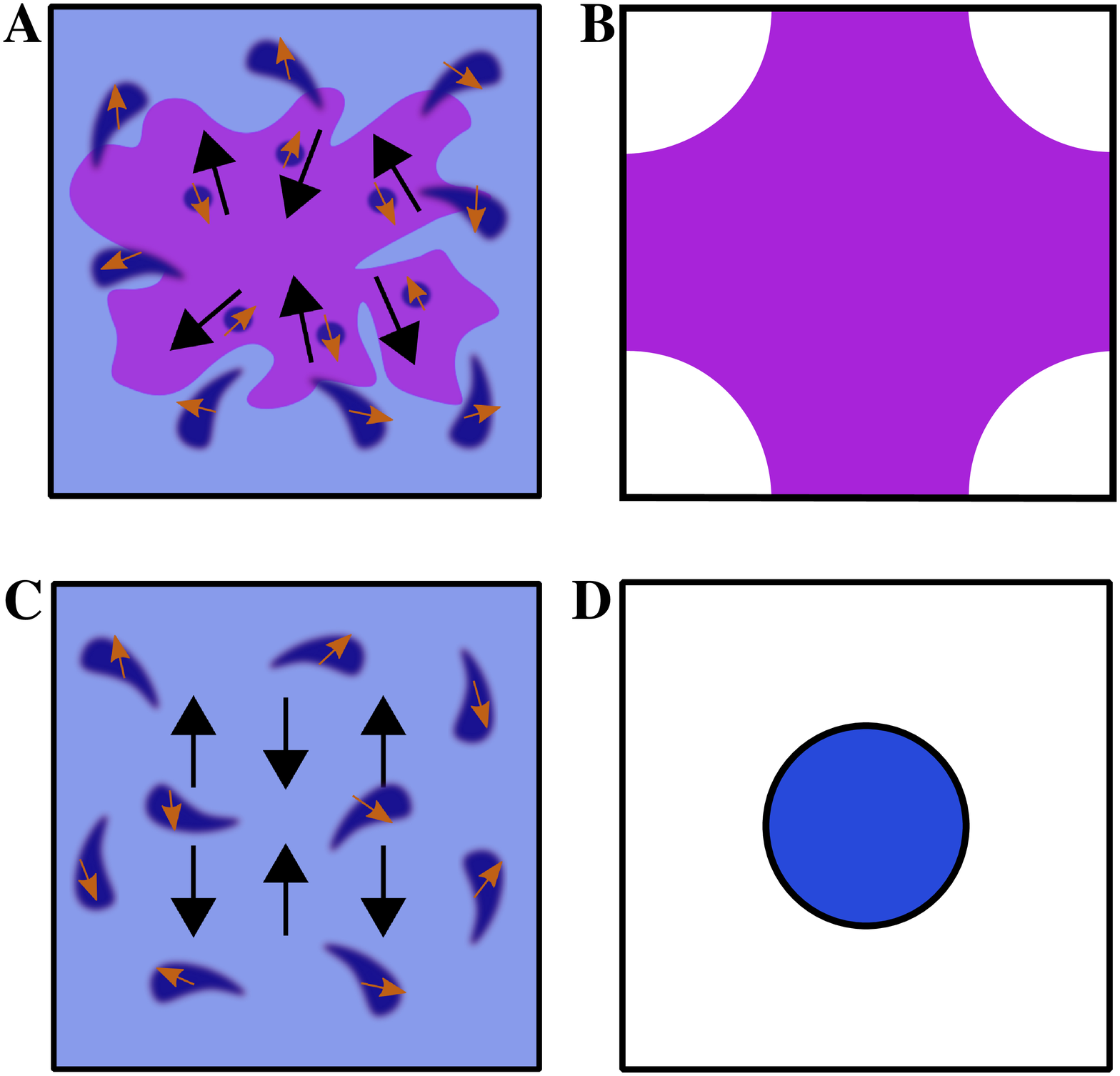}}
\centerline{} \centerline{} 
\centerline{\Large Figure~2}
\end{figure}

\newpage

\begin{figure}[h!]
%%%\centerline{\includegraphics[width=0.8\linewidth]{Figure3_version7.eps}}
%centerline{\includegraphics[width=0.6\linewidth]{Fig3a-scaling_ccau01_review.eps}}
%\vskip 0.5cm
%\centerline{\includegraphics[width=0.4\linewidth]{Fig3b-Fig.8a.eps}}
\centerline{\includegraphics[width=1.0\linewidth]{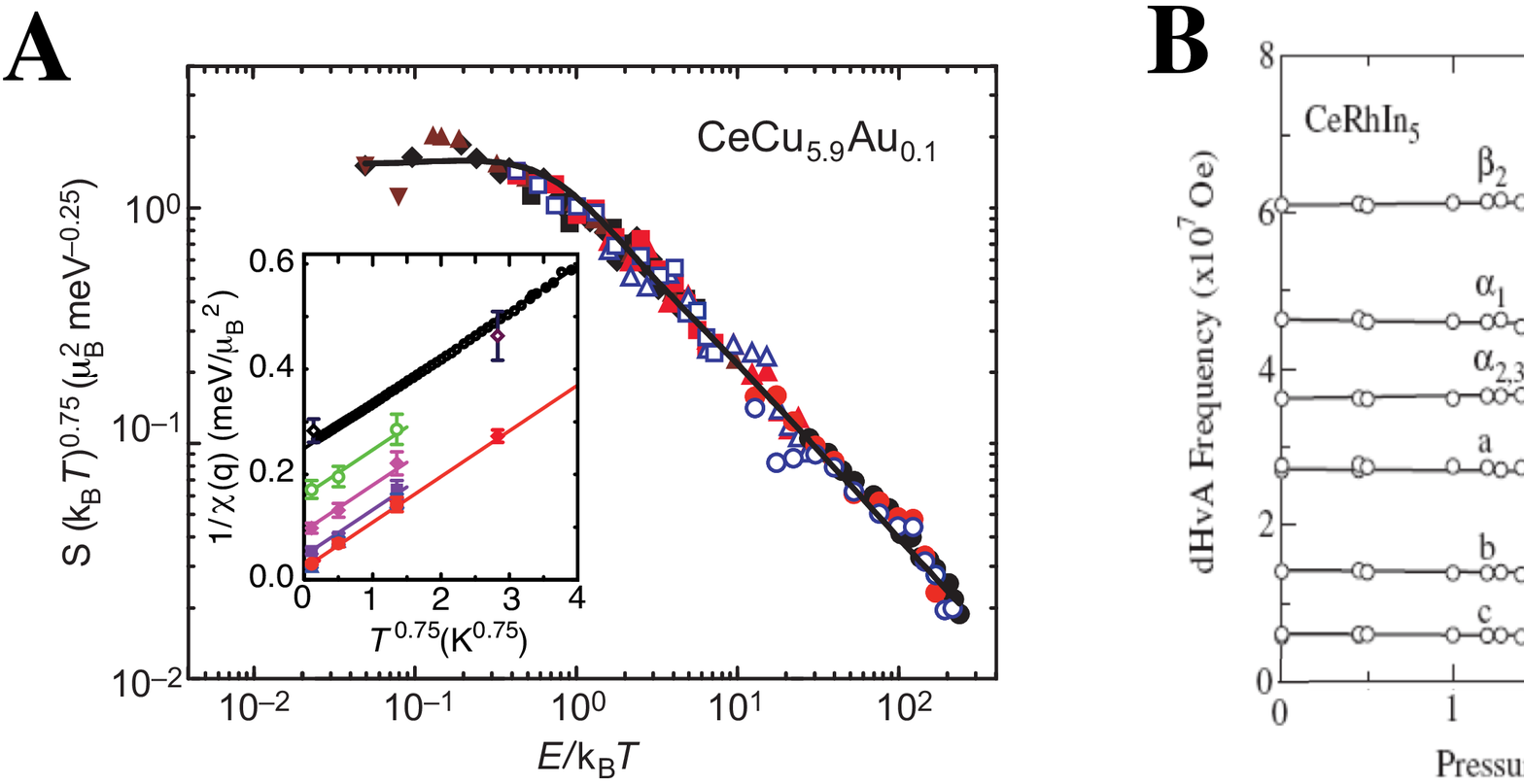}}
\centerline{} \centerline{} 
\centerline{\Large Figure~3}
\end{figure}

\newpage

\begin{figure}[h!]
%%%\centerline{\includegraphics[width=0.8\linewidth]{Figure4_version5.eps}}
%%\centerline{\includegraphics[width=0.6\linewidth]{Fig4a-Science_Phasendiagramm.eps}}
%%\centerline{\includegraphics[width=0.6\linewidth]{Fig4b-FWHM.eps}}
%\centerline{\psfig{figure=Fig4a-Science_Phasendiagramm.eps,width=0.4\linewidth,angle=0}
%\hskip 0.2 cm
%\psfig{figure=Fig4b-FWHM.eps,width=0.4\linewidth}}
%%\centerline{\includegraphics[width=0.6\linewidth]{Fig4c-PD_17Ir.eps}}
%%\centerline{\includegraphics[width=0.6\linewidth]{Fig4d-PD_2.5Ir.eps}}
%\vskip 0.5cm
%\centerline{\psfig{figure=Fig4c-PD_17Ir.eps,width=0.4\linewidth,angle=0}
%\hskip 0.2 cm
%\psfig{figure=Fig4d-PD_2.5Ir.eps,width=0.4\linewidth}}
%%\centerline{\includegraphics[width=0.6\linewidth]{Fig4e-PD_3Co.eps}}
%%\centerline{\includegraphics[width=0.6\linewidth]{Fig4f-PD_at_T_eq_0.eps}}
%\vskip 0.5 cm
%\centerline{\psfig{figure=Fig4e-PD_3Co.eps,width=0.4\linewidth,angle=0}
%\hskip 0.2 cm
%\psfig{figure=Fig4f-PD_at_T_eq_0.eps,width=0.4\linewidth}}
\centerline{\includegraphics[width=0.9\linewidth]{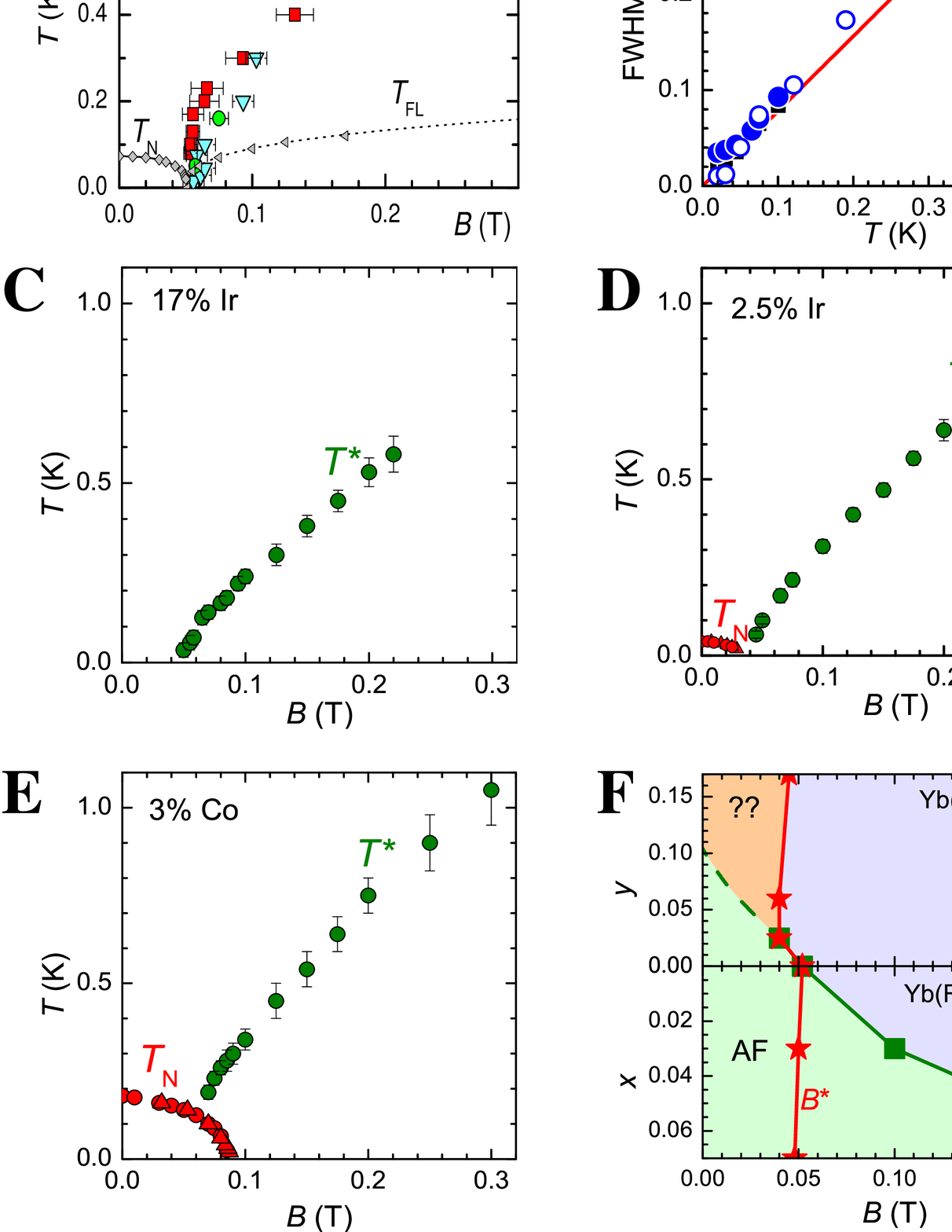}}
%\centerline{}
\centerline{}
\centerline{\Large Figure~4}
\end{figure}

%\newpage
%
%\begin{figure}[h!]
%%\centerline{\includegraphics[width=0.6\linewidth]{Fig4e-PD_3Co.eps}}
%%\centerline{\includegraphics[width=0.6\linewidth]{Fig4f-PD_at_T_eq_0.eps}}
%\centerline{\psfig{figure=Fig4e-PD_3Co.eps,width=0.4\linewidth,angle=0}
%\hskip 0.2 cm
%\psfig{figure=Fig4f-PD_at_T_eq_0.eps,width=0.4\linewidth}}
%\centerline{}
%\centerline{}
%\centerline{\Large Figure~4}
%\end{figure}

\end{document}